\documentclass[lettersize,journal]{IEEEtran}
\usepackage{amsmath,amsfonts}
\usepackage{array}
\usepackage[caption=false,font=normalsize,labelfont=sf,textfont=sf]{subfig}
\usepackage{textcomp}
\usepackage{stfloats}
\usepackage{url}
\usepackage{verbatim}
\usepackage{graphicx}
\usepackage{cite}
\usepackage{xcolor}
\usepackage{booktabs}
\usepackage{multirow}
\usepackage{algorithm}
\usepackage{algpseudocode}
\usepackage{hyperref}
 
\begin{document}
\title{Pest Manager: A Systematic Framework for\\ Precise Pest Counting and Identification in\\ Invisible Grain Pile Storage Environment}

\author{Chuanyang Ma, Jiangtao Li, Xingqun Qi, Muyi Sun, and Huiling Zhou
\vspace{-8mm}
\thanks{This research was partially funded by the National Natural Science Foundation of China (Grant No. 62306309) and the Postdoctoral Fellowship Program of China (Grant No. GZC20240162). Corresponding Author: Muyi Sun.}
\thanks{Chuanyang Ma, Jiangtao Li, Muyi Sun, and Huiling Zhou are with the
School of Artificial Intelligence, Beijing University of Posts and Telecommunications. 
(e-mail: chuanyang.ma@bupt.edu.cn, lijiangtao@bupt.edu.cn, muyi.sun@bupt.edu.cn, huiling@bupt.edu.cn). }
\thanks{Xingqun Qi is with Academy of Interdisciplinary Studies, The Hong Kong University of Science and Technology. (e-mail: xingqun.qi@connect.ust.hk). }
}

\maketitle 

\begin{abstract}
Pest infestations pose a significant risk to both the quality and quantity of stored grain, resulting in substantial economic losses.
Precise and prompt pest monitoring is crucial for mitigating these losses.
Traditional methods rely heavily on manual sampling strategies, which are time-consuming, labor-intensive, and often lack the accuracy for effective monitoring.
Recently, several intelligent monitoring methods have been introduced. 
However, they focus on the visible grain pile surface primarily and lack the ability to penetrate deeper inside the pile.
In this paper, we introduce a systematic framework: \textcolor{blue}{Pest Manager} for precise pest counting and identification within the invisible grain pile environment.
The framework consists of three components: an improved grain probe trap \textcolor{blue}{PestMoni}, a pest drop dataset \textcolor{blue}{PestSet} collected by PestMoni, and a multi-task Transformer-based architecture \textcolor{blue}{PestFormer} for pest counting and identification.
Specifically, PestMoni utilizes two pairs of infrared diodes and photodiodes, enabling precise record of pest drops through an asymmetric orthogonal diode layout with a tailored circuit. 
Based on PestMoni, we develop PestSet, an infrared-perception signal dataset for pest drops, which is then used for pest counting and species identification of five common grain storage pests.
Furthermore, we propose PestFormer, a multi-task pest analysis model with a conditional modification module, which can reduce deviations across different PestMonis and thereby normalize the data processing workflow. 
Consequently, extensive experiments have been conducted to verify the rationality of PestMoni design, the availability of the collected PestSet, and the effectiveness of PestFormer. 
Finally, accompanied by stable sensing capabilities, our PestFormer achieves State-Of-The-Art results in pest counting and identification with the accuracy of 99.2\%  and 86.9\%, which demonstrates great potential for pest management in invisible storage environment.
\end{abstract}


\section{Introduction}
\IEEEPARstart{P}{ost-harvest} food loss is a critical global issue, with over one-quarter of food lost or wasted during post-harvest processes in certain regions. 
Among all agricultural products, grains experience the most substantial losses \cite{b1}.
These losses span various stages from farm to consumer, including harvesting, threshing, cleaning, drying, storage, processing, and transportation. 
Notably, the storage phase alone accounts for about 55\% of the total post-harvest losses \cite{b2}. 
A primary cause is pest infestation, particularly in developing regions due to inadequate storage facilities and lack of advanced pest control systems \cite{b1,b2}.
Implementing precise and efficient pest monitoring can significantly safeguard stored grain, enhance food availability, alleviate resource constraints, and contribute to reducing hunger.

\begin{figure}[!t]
\centerline{\includegraphics[width=0.95\columnwidth]{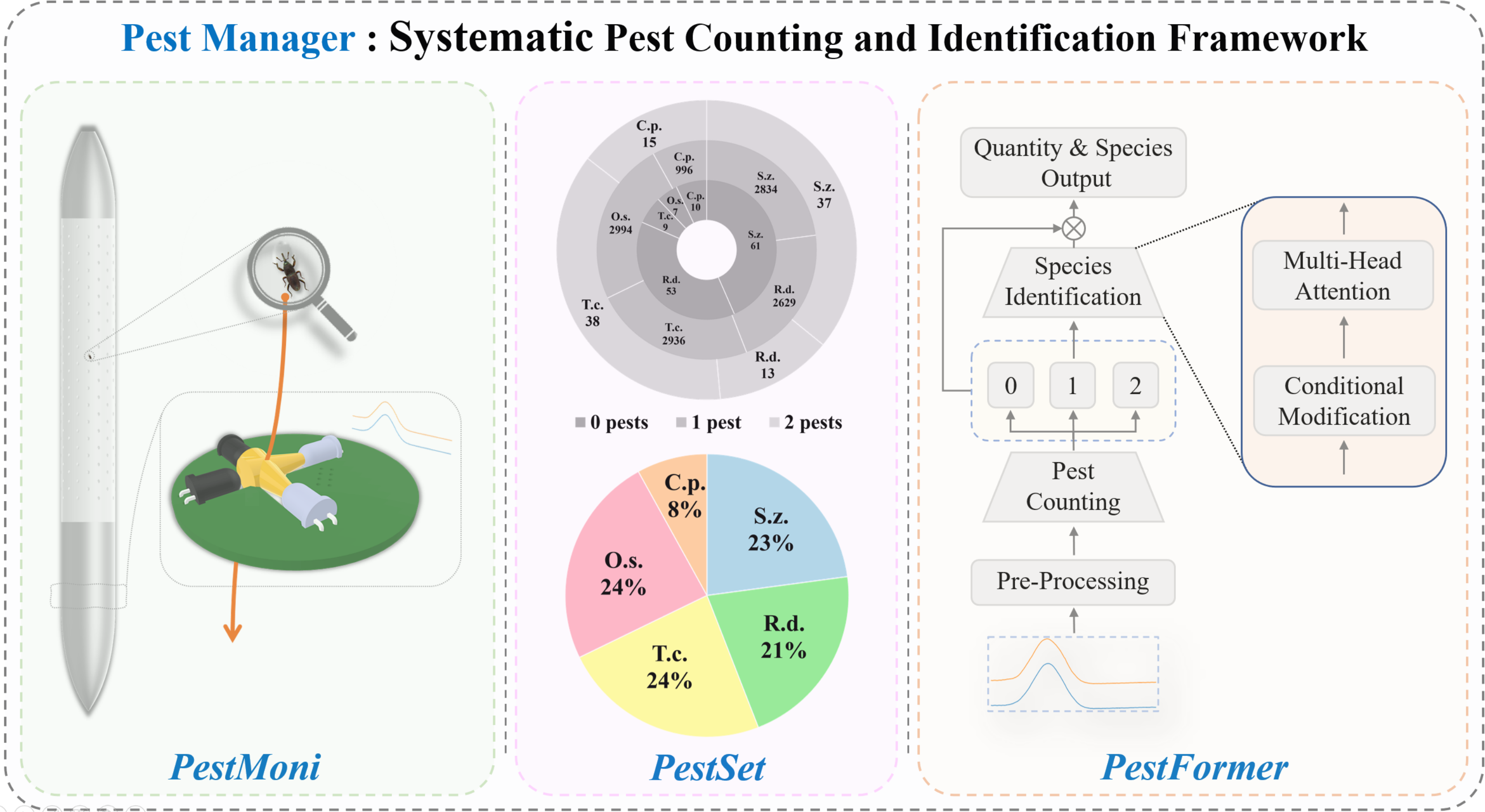}}
\caption{Overview of Pest Manager Framework. 
PestMoni: asymmetric orthogonal infrared perception. 
PestSet: multi-class pest drop dataset (Top: 3-scale pest quantities within a single perception data; Bottom: 5-species distribution).
PestFormer: multi-task pest analysis.}
\vspace{-4mm}
\label{fig1}
\end{figure}

Generally, grain pest monitoring techniques are primarily divided into two categories: \textbf{visible environment surveillance on the grain surface} and \textbf{invisible environment monitoring inside the grain pile}.
\textbf{On the surface}, traditional methods such as visual inspection are straightforward and simple, but they only provide qualitative assessments and rely heavily on subjective judgments, leading to potential inaccuracies and inconsistencies \cite{b2,b3}. 
In recent years, grain storage infrastructure has been enhanced by integrating high-definition cameras at numerous facilities.
Utilizing these cameras, researchers have developed efficient image-based pest detection methods that rapidly and precisely identify pests on the surfaces of grain piles \cite{b4,b5}.
Unlike surface-level monitoring, pests \textbf{within the invisible environment of grain piles} are more concealed and cannot be directly observed, necessitating a sensitive and precise insertion-based monitoring method.
Despite the low cost, flexibility, and intuitiveness associated with manual sampling using grain triers and sieves, it still presents disadvantages such as being labor-intensive and limited in efficiency when monitoring pests within grain piles \cite{b2,b3,b4}.
In addition, grain probe traps, another commonly used device for pest monitoring in grain piles, were initially developed to facilitate continuous pest monitoring and capture. 
These traps are widely adopted due to their effectiveness and user-friendly design, yet they require routine inspections and cleaning, which increase labor demands and may lead to delays in the data collection process \cite{b3,b4}. 
Furthermore, the frequency and timeliness of these inspections critically impact the accuracy and reliability of the monitoring data.
Expanding upon traditional designs, researchers have upgraded conventional grain probe traps by integrating specialized electronic circuitry for automatic triggering and recording \cite{b6,b7}. 
These upgraded grain probe traps excel at pest counting but often struggle with precise species identification, typically requiring manual checks for accurate taxonomy \cite{b3}. 
To address these challenges, we introduce PestMoni, an improved electronic grain probe trap designed for both pest counting and identification within invisible grain pile environment.
Notably, PestMoni provides the capability to record the complete waveform generated by pests shading infrared light during their drops. 
This waveform contains distinguishable features make species identification accessible through time series analysis methods.

To effectively apply time series classification techniques, a robust and representative dataset is indispensable. 
In a previous study, Shuman et al. \cite{b8} conducted data collection experiments on four grain storage pests using their designed electronic grain probe traps. 
However, the pests they used in their experiments were deceased, which failed to accurately replicate the actual operational conditions of the monitoring devices.
Additionally, their methodology focused solely on selected waveform features such as peaks, durations, and intervals, resulting in the loss of valuable continuous sequence features.
Currently, there is a lack of a comprehensive pest drop dataset based on electronic grain probe traps.
To address these deficiency, we have developed PestSet, leveraging PestMoni to capture the complete waveform of drops for five common live pests in the grain storage industry.
By providing a detailed record of the changes in infrared light intensity shaded by pests during the drop process, PestSet establishes the groundwork for accurate counting and species identification.

Building on the aforementioned techniques, time series classification methods are essential for analyzing pest waveforms.
Time Series Classification (TSC) is a classic problem in data mining and machine learning, approached through three main methods: 1) Traditional feature-based, 2) Deep learning-based, and 3) Transformer-based methods.
Traditional methods extract data features such as statistical properties, frequency domain features, and morphological aspects \cite{b9, yan2017fast}.
However, these features can not capture all relevant information, especially in complex datasets, and require extensive time and expertise for feature selection.
Deep learning methods utilize architectures like Recurrent Neural Networks (RNNs) \cite{b10,b11,b12} and Convolutional Neural Networks (CNNs) \cite{b13,b14} to automatically extract features, reducing the need for manual intervention.
However, these methods require large annotated datasets and significant computational resources for effective training.
Beneficial for the ability to handle sequential data through attention mechanisms, Transformer-based methods allow for the capture of long-range dependencies and also maintain the sequence order using positional encoding, which is essential for precise time series classification \cite{b15,b16}.
Leveraging the merits of Transformer, we introduce PestFormer, a condition-modified Transformer for multi-task pest waveform analysis. 
By incorporating a device-aware conditional modification module, PestFormer can extract pest waveform representations that are independent of individual devices, enabling accurate pest counting and species identification across various PestMoni data sources.

To sum up, in this paper, we introduce a systematic framework \textbf{Pest Manager}, to enhance grain storage pest monitoring within the invisible environments of grain piles, providing accurate pest counts and species identification. 
This framework comprises an improved electronic grain probe trap PestMoni, a comprehensive pest drop dataset PestSet (derived from PestMoni), and a Transformer-based architecture PestFormer for pest waveform analysis, as shown in Fig. 1. 
Specifically, we enhanced PestMoni using two pairs of infrared diodes and photodiodes arranged in an asymmetric orthogonal layout.
Combined with a tailored circuit, this configuration ensures precise pest counting and facilitates the complete acquisition of time series data during the pest drop process, capturing distinct pest characteristics.
Subsequently, we conducted extensive data collection on five common live pest species: \textit{Sitophilus zeamais} (S. z.), \textit{Rhizopertha dominica} (R. d.), \textit{Tribolium castaneum} (T. c.), \textit{Oryzaephilus surinamensis} (O. s.), and \textit{Cryptolestes pusillus} (C. p.).
This effort resulted in PestSet, comprising 12,632 pest drop waveforms, which are utilized for precise pest counting and species identification.
Additionally, we designed PestFormer with a conditional modification module for multi-task pest analysis, which reduces the deviations caused by different PestMonis and standardizes the data processing workflow.
After extensive experimental validation, our PestFormer has demonstrated exceptional performance, achieving a remarkable accuracy of 99.2\% for pest counting and 86.9\% for species identification.
In summary, the key contributions of this work are as follows:
\begin{itemize}
    \item \textbf{Systematic Framework:} We introduce \textcolor{blue}{\textbf{Pest Manager}}, an intelligent system for automated pest counting and identification in invisible grain pile storage environment.
    
    \item \textbf{Pest Monitor Design:} We propose an improved grain probe trap \textcolor{blue}{\textbf{PestMoni}}, with an asymmetric orthogonal diode layout and a tailored circuit to record pest drops for pest counting and species identification.
    
    \item \textbf{Pest Dataset Collection:} We develop \textcolor{blue}{\textbf{PestSet}}, a pest drop dataset based on infrared signals, comprising 12,632 waveforms of five common grain storage pests, facilitating the pest counting and species identification.

    \item \textbf{Transformer-Based Architecture:} We present \textcolor{blue}{\textbf{PestFormer}}, a condition-modified Transformer for multi-task pest analysis, achieving high-precision time series classification associated with device-sensitive data.
    
    \item \textbf{Experimental Validation:} Extensive experiments have been conducted to comprehensively demonstrate the rationality of PestMoni, the availability of PestSet, and the effectiveness of PestFormer, which excels in state-of-the-art pest counting and species identification.
\end{itemize}

\section{RELATED WORK}
\label{sec:related work}
In this section, we introduce previous research related to grain pile pest monitors, pest drop datasets and time series classification methods, which are also the main components of the research framework in this paper.

\begin{figure*}[!t]
\centering
\includegraphics[width=0.95\textwidth]{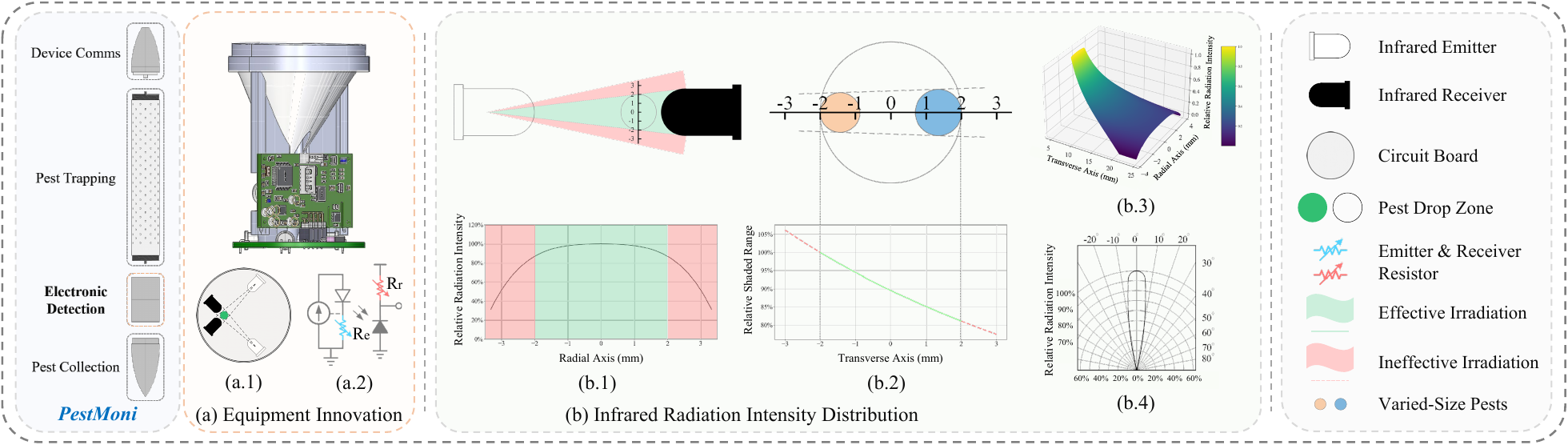}
\caption{Structural Design in \textcolor{blue}{\textit{\textbf{PestMoni}}}: Device Comms, Pest Trapping, Electronic Detection and Pest Collection. 
(a) Core innovations in Electronic Detection.
(a.1) Asymmetric orthogonal layout of infrared emitter and receiver; 
(a.2) Simplified diagram of the tailored circuit for enhanced perception.
(b) The distribution of infrared radiation intensity.
(b.1) Relative radiation intensity analysis on radial; 
(b.2) Relative shaded range analysis on transverse; 
(b.3) Relative radiation intensity distribution across orthogonal biaxial plane; 
(b.4) Radiation characteristics of infrared diodes \cite{b17}.}
\label{fig2}
\end{figure*}

\subsection{Grain Pest Monitor and Dataset}
Grain pest monitoring is the initial step in pest control. 
In recent years, grain probe traps have emerged as a widely adopted type of grain pest monitor \cite{b4}. 
These traps commonly feature perforated cylindrical tubes as trapping components, leveraging the burrowing behavior of pests to attract them into the bottom collection area through these apertures \cite{b4,b6}.
Hagstrum et al. \cite{b18} reveal that these traps detect pests 37 days earlier than typical grain trier samples. 
However, traditional grain probe traps require frequent manual inspections and cleaning, significantly increasing labor demands.
To overcome this limitation, Shuman et al. \cite{b6} introduce the Electronic Grain Probe Insect Counter (EGPIC), which automates the analysis of falling pests, thereby reducing the necessity for manual inspections.
The EGPIC employs a symmetrical pair of infrared diodes and phototransistors as its sensing elements.
Dropping pests can shade the infrared light emitted by the diodes, resulting in a voltage change in the phototransistors and the generation of pulse signals.
These pulses are then captured for pest counting \cite{b6}.
Subsequently, a dual infrared-beam version is introduced to enhance pest detection capability, incorporating the collection of pulse characteristics such as signal peak amplitudes and durations \cite{b8}.
Through the analysis of these characteristics, they successfully differentiate flat grain beetles from red flour beetles or rice weevils.
However, their collected data is derived from deceased pests and are not suitable for real-world applications.
Additionally, the stability of perception in these monitors remains limited, and the acquisition and utilization of pest drop data are insufficient.
Inspired by the aforementioned research, we developed PestMoni, a perception-stable monitor with an asymmetric orthogonal diode layout designed for real-time monitoring of live pests in storage environments and complete collection of pest drop waveforms.

\subsection{Time-Series Classification}
As a key task in time series analysis, numerous time series classification methods have been derived from classical tools \cite{b9}, such as TSF \cite{b19}, DWT \cite{b20} and Shapelets \cite{b21}.
However, real-world time series variations often present complexities that surpass pre-defined patterns, thereby limiting the practical applicability of these classical methods.
Recently, TimesNet \cite{b22} introduces an efficient framework for time series analysis, leveraging advancements in computer vision to transform time series data into two-dimensional representations akin to image data, achieving promising results.
Additionally, Transformer methods based on the self-attention mechanism have demonstrated formidable capabilities in sequential data processing \cite{b15,b16}.
The multi-head attention mechanism further enhances learning by identifying temporal dependencies among different time points \cite{b23,b24,b25}.
In addition, benefiting from the positional encoding operation, they excel at capturing the sequential information within time series data.
Specifically, Zerveas et al. \cite{b26} implemented a learnable positional embedding layer within the Transformer, achieving more advanced results than vanilla positional encoding by jointly learning with other model parameters.
However, the above-mentioned methods have mainly focused on improving the effectiveness of standardized data analysis and have not addressed the accuracy degradation associated with device-sensitive data.
In this paper, we introduce a novel Transformer architecture called PestFormer for analyzing pest drop waveforms, which incorporates conditional modification capabilities to mitigate the device-aware performance decline.

\vspace{2mm}

\section{Methodology}
In this section, we provide a detailed overview of the three core components of the \textbf{Pest Manager} framework: the design of PestMoni, the construction of PestSet, and the architecture of PestFormer. 
Through the integrated application of these components, we demonstrate how Pest Manager effectively monitors and identifies pests in invisible storage environment.

\subsection{PestMoni Design}
PestMoni is composed of four parts: Device Comms, Pest Trapping, Electronic Detection, and Pest Collection, as illustrated in Fig. 2. 
Grain storage pests first enter PestMoni through Pest Trapping, where they climb and linger on the inner walls.
Due to gravity, these pests eventually fall. 
As the pests pass through Electronic Detection, they shade the infrared light, generating voltage change waveforms.
These waveforms are then processed and transmitted to the server via Device Comms. Finally, the pests fall into Pest Collection.

Pest Trapping consists of a hollow cylinder (50mm outer diameter, 44mm inner diameter, 280mm length) with uniformly spaced 2mm bait holes angled upwards at 45° from the outer to the inner surface.
This design leverages pests' burrowing behavior to enhance trapping efficiency while reducing grain debris entry.
Electronic Detection comprises a collection funnel and two circuit boards, as detailed in Fig. 2(a).
The funnel has a 4mm hole at its base, extending downward for 20mm to form a narrow passage that confines the Pest Drop Zone. 
It is coated with polytetrafluoroethylene to prevent blockages caused by pests.
The two circuit boards are designated for control and detection functions, respectively. 
The control circuit runs the program, processes data, and transmits data, while the detection circuit specifically perceives pest drops.
Device Comms acts as a crucial link for transmitting power and data from the server via connections to the gateway. 
Pest Collection is designed as a hollow cone with a smooth interior to trap pests and prevent them from returning to the detection zone, thus reducing the risk of false readings and system malfunctions.

\textbf{Core innovations in Electronic Detection for PestMoni.} 
PestMoni features an asymmetric orthogonal layout of infrared components and a tailored circuit. 
Infrared diodes convert electricity into infrared radiation, which is then received by photodiodes, forming a pair perception sensor.
PestMoni utilizes a 5mm infrared diode with a 10-degree half-power angle, generating a fan-shaped radiation pattern that decreases in intensity over distance and angle, refer to Fig. 2(b.4).
Positioned opposite this diode is a corresponding photodiode, aligned with the center of the Pest Drop Zone to capture the emitted light.
This center is defined as the origin, where the line passing through the diode centers constitutes the transverse axis, and the radial axis runs perpendicular to it.

\begin{table}[htbp]
    \centering
    \caption{Relative Radiation Intensity at Different Sampling Positions}
    \begin{tabular}{cc}
        \toprule
        \textbf{Radial Axis (mm)} & \textbf{Relative Radiation Intensity (\%)} \\
        \midrule
        $0$ & $100$ \\
        $1$ & $97.96$ \\
        $1.5$ & $95.19$ \\
        $2$ & $86.45$ \\
        $3$ & $51.74$ \\
        \bottomrule
    \end{tabular}
\end{table}
Data sampled from Fig. 2(b.4) in Table I is used to construct a function that maps radial axis coordinates to relative radiation intensity, as follows:
\begin{align}
RRI = &\, 0.0027r^6 + 4 \times 10^{-12}r^5 - 0.4593r^4 + 2 \times 10^{-10}r^3\nonumber \\
      &\, - 1.4844r^2 + 3 \times 10^{-9}r + 100.11
\end{align}
where $RRI$ represents relative radiant intensity, and $r$ denotes radial axis coordinates.
Fig. 2(b.1), plotted according to (1), illustrates the attenuation of $RRI$ along the radial axis, using green to indicate effective light and red for areas outside the receiver's reach.
Additionally, the shaded area along the transverse axis varies with the object's position.
Fig. 2(b.2) shows the same shaded range caused by pests of different sizes, represented by multi-colored circles.
\textbf{To quantify this variation}, we utilized a 1mm radius circle, centered at -2mm on the transverse axis, as a reference for the shaded range, and then modeled the relative shaded range at different positions, represented as follows:
\begin{align}
RSR = \frac{\theta}{\theta_{ref}} = \frac{\arctan(\frac{1}{D+t})}{\arctan(\frac{1}{D-R})}
\end{align}
where $RSR$ represents the relative shaded range, $\theta$ denotes the actual angular extent shaded by the pest, while $\theta_{ref}$ corresponds to the reference shaded angular extent.
The distance $D$ from the origin to the light source is set at 19.13mm.
$R$ signifies the 2mm radius of Pest Drop Zone, and $t$ indicates the transverse axis coordinate.
Figure 2(b.2) shows that as the coordinates increase along the transverse axis, $RSR$ decreases (depicted by the green line).
To address the issue of similar shading at specific positions for pests of varying sizes, precise localization during their descent is crucial.
In Fig. 2(b.3), the radiation intensity distribution on the orthogonal biaxial plane reveals a symmetrical pattern across the transverse axis, making precise localization challenging.
In light of this, we have added an additional pair of sensors. 
Positioned orthogonally to the first sensor, this configuration allows for overlapping perceptions of pest drops. 
\textbf{By integrating the data from both sensor pairs, we can locate pests and enhance the sensing dimensions of PestMoni.} 

\begin{figure}[t]
\centerline{\includegraphics[width=0.95\columnwidth]{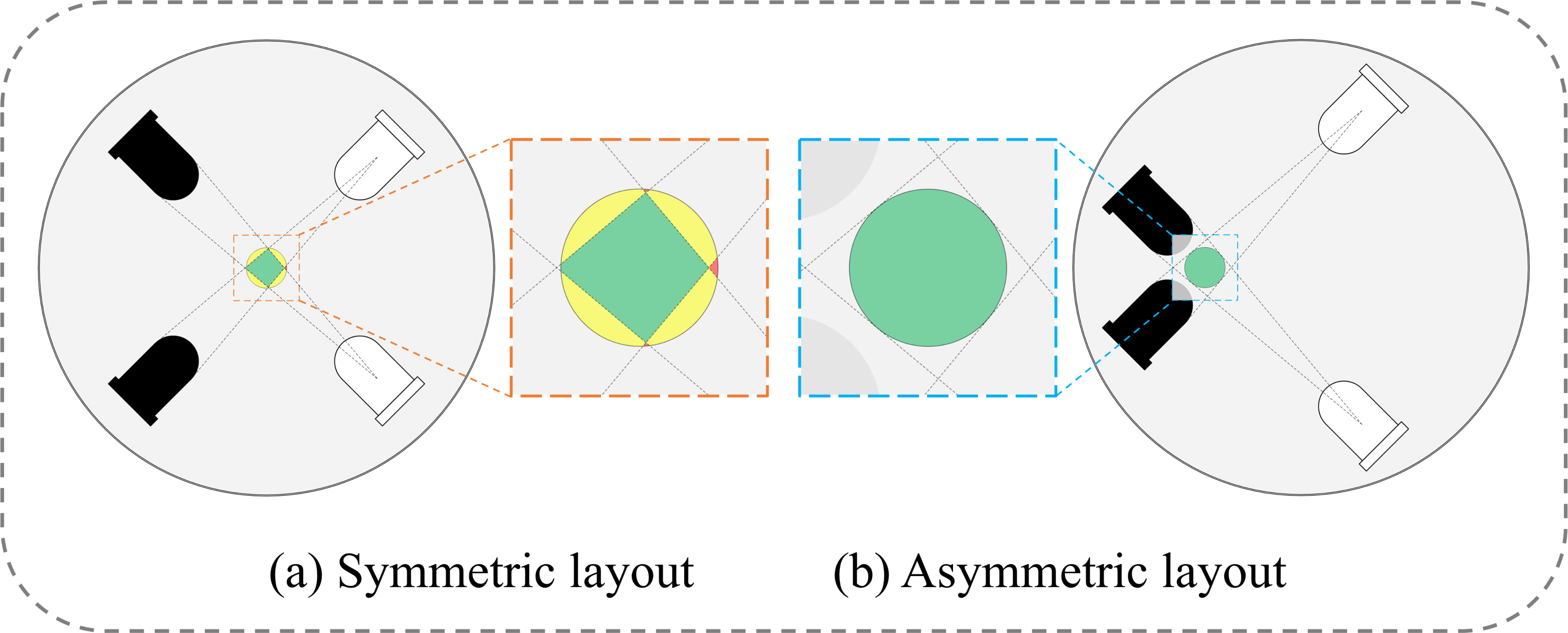}}
\caption{Comparison analysis of perception range between symmetric (left) and asymmetric (right) layout.}
\label{fig3}
\end{figure}

To avoid confusion, it's crucial to clarify two distinct concepts: the radiation angle of the infrared diodes and the effective reception angle of the photodiodes.
In the context of perceiving pest drops, the latter is the critical parameter, while the former only needs to surpass it.
Shuman et al. \cite{b6} employ a symmetric layout that introduces significant challenges as illustrated in Fig. 3(a). 
The gray dashed line indicates the maximum effective range of the receivers, dividing the central Pest Drop Zone into several areas: green represents the normal perceiving area, yellow denotes regions where only one pair of sensors is effective, and red highlights the perceiving blind spots.
\textbf{To resolve these issues and extend sensor coverage for perception}, we introduce an asymmetric layout as shown in Fig. 3(b). 
This configuration ensures that the entire Pest Drop Zone falls within the detection range of both sensor pairs, thereby eliminating the perceiving blind spots.

\begin{figure*}[!t]
\centering
\includegraphics[width=\textwidth]{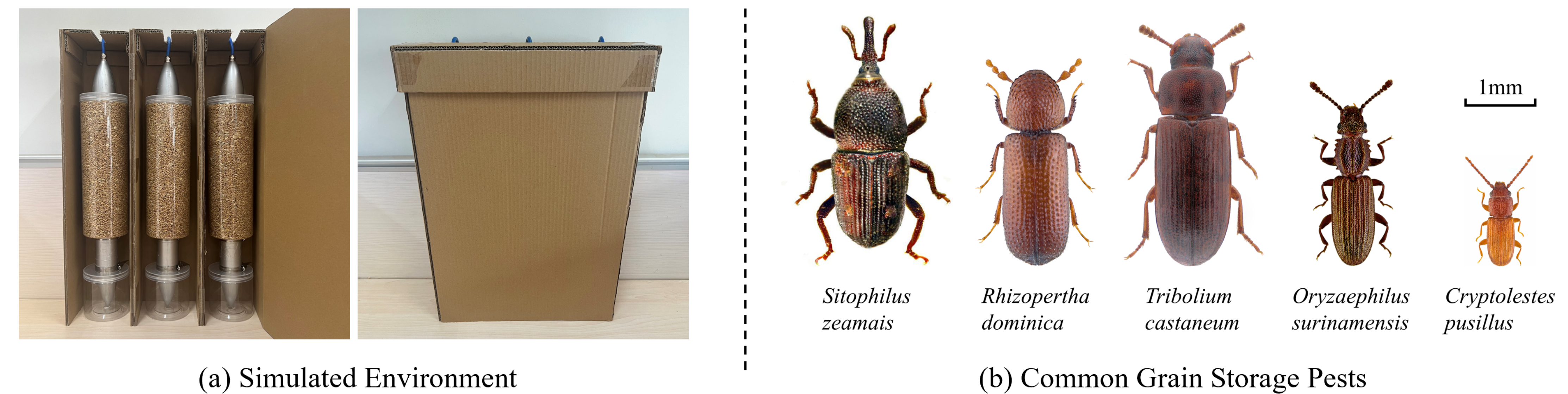}
\caption{Establishment of PestSet: 
(a) A simulated invisible grain pile environment in the laboratory for pest perception data collection.
(b) Approximate body lengths of different grain storage pests: Sitophilus zeamais (\~3.6 mm), Rhizopertha dominica (\~3 mm), Tribolium castaneum (\~3.5 mm), Oryzaephilus surinamensis (\~2.7 mm), Cryptolestes pusillus (\~1.6 mm) (Source: Wikipedia, \href{https://commons.wikimedia.org/wiki/}{https://commons.wikimedia.org/wiki/}).}
\label{fig4}
\end{figure*}

Measuring voltage changes across the receiver’s terminals with an Analog-to-Digital Converter (ADC) enables the perception of pest drops.
However, this method inevitably leads to quantization errors.
\textbf{To minimize the impact of these errors, we tailor the driving circuit}, as depicted in Fig. 2(a.2).
The circuit's relationships are expressed by the following equation:

\begin{gather}
E_{e} = k_{1}I_{e} + c_{1} = k_{1}\frac{a}{R_{e}} + c_{1} \label{eq3} \\
U_{r} = VCC - R_{r}I_{r} = VCC - R_{r}\left ( k_{2}E_{r} + c_{2} \right ) \label{eq4} \\
\rho \left ( E_{r}, E_{e} \right ) > 0 \label{eq5}
\end{gather}
where \(E_e\) and \(E_r\) represent the total radiation intensity emitted by the emitter and the total light intensity received by the receiver, respectively.
\(I_e\) and \(I_r\) indicate the currents in the emitting and receiving circuits, while \(U_r\) is the voltage difference across the terminals of the receiver.
\(VCC\) is the positive voltage in the receiving circuit, and $\rho$ denotes the Pearson correlation coefficient. 
The system parameters \(k_1\), \(k_2\), \(c_1\), \(c_2\) and \(a\) are constants determined by the component characteristics.
Both coefficients \(k_1\) and \(k_2\), as well as the parameter \(a\), are positive.

With the same ADC resolution, to reduce the signal's quantization error, it is necessary to amplify the voltage change caused by pest dropping through the Electronic Detection, denoted as $\Delta U_r$. 
This can be achieved by appropriately increasing the receiver resistance $R_r$ and decreasing the emitter resistance $R_e$.
First, we analyze the receiving circuit.
Assuming that the radiation from the infrared diode and the light shaded by pests remain constant, this results in a fixed current in the receiving circuit. 
Using a larger $R_r$ will amplify the voltage change across the receiving resistor. 
Since the sum of the voltage drops across the receiver resistor and the photodiode remains constant at VCC, this amplifies the voltage change across the photodiode, thus achieving the intended purpose.
However, \( R_r \) cannot be increased indefinitely. 
When the photodiode operates in a reverse bias state, its PN junction generates junction capacitance, which increases as the reverse bias voltage decreases. 
If \( R_r \) becomes very large, it results in a significant voltage drop across the resistor, reducing the voltage difference across the photodiode and consequently increasing the junction capacitance.

\begin{equation}
\tau = R \times C
\end{equation}

\begin{equation}
V(t) = V_{\text{max}} \left(1 - e^{-t/\tau}\right)
\end{equation}

Equation (6) shows that larger load resistance and junction capacitance lead to a larger time constant. 
Consequently, as described in (7), an increased time constant further extends the system's response time.
This will be detrimental to a quick response to pest drops.

For the emitting circuit, reducing $R_e$ can increase the current flowing through the infrared diode, thereby increasing the intensity of infrared radiation. 
This results in more infrared light being shaded when pests enter the detection area.
However, $R_e$ should not be too small either, as excessively high radiation intensity can cause the photodiode to receive high levels of radiation even when pests are blocking it. 
This can potentially lead to a significant voltage drop across \( R_r \), rendering the photodiode's operating voltage insufficient for its linear region.
As a result, it may fail to detect pest drops, leading to the possibility of smaller pests being ignored.

Therefore, when selecting specific components, we first choose the appropriate receiver resistor based on the actual application scenario, ensuring that the response time of the receiving circuit is within an acceptable range. 
Then, we adjust the value of the emitter resistor to ensure that the photodiode operates in its linear region, \textbf{thereby increasing $\Delta U_r$, further mitigating the impact of quantization errors and enhancing the sensor's perception.}

\subsection{PestSet Collection}
To construct a comprehensive dataset, we utilized three PestMonis for data acquisition and set up the configuration shown in Fig. 4(a) to simulate an invisible grain pile storage environment.
Transparent plastic cylinders, each with a diameter of 10 cm and a height of 30 cm, were used for the collection process.
These cylinders were equipped with matching holes specifically designed for PestMoni to accommodate its Pest Trapping.
The containers were filled with wheat, and live pests were introduced in batches.
The captured pests were collected in the Pest Collection unit for reuse.

We set the ADC resolution to 12 bits, providing 4096 quantization levels to ensure high-precision sampling. 
The scanning mode is enabled to continuously sample two channels, with data transferred via Direct Memory Access (DMA) to a 512-byte buffer, allowing a total of 256 data points to be stored (128 data points per channel, with each data point occupying 2 bytes).
When the buffer overflows, the pointer resets to the start of the buffer, overwriting the old data.
The ADC’s external trigger source is set to the timer update event, and the continuous conversion mode is disabled to control the sampling frequency via the timer period.
We set the timer period to 200 $\mu$s, resulting in a 5 kHz sampling frequency. 
The buffer length covers 12.8 ms of signal, which is sufficient for monitoring most normal pest drops in practical applications.

\begin{algorithm}[H]
\caption{\textbf{TransferData}: Transfer data from the DMA buffer to the waveform list.}
\begin{algorithmic}[1]
\Require Start Position $i \in [0, L-1)$, End Position $j \in (0, L-1]$ such that $0 \leq i < j \leq L-1$, DMA Buffer \( \text{ADCBuf} \in \mathbb{R}^{L} \), Waveform List \( \text{WaveList} \in \mathbb{R}^{M \times L} \), Write Index \( m \in [0,  M-1]\)
\For{$l \gets i \text{ to } j \text{ step } 2$}
    \State $\text{WaveList}[\text{\( m \)}][(L - j + l - i) / 2] \gets \text{ADCBuf}[l]$
    \State $\text{WaveList}[\text{\( m \)}][(2 \times L - j + l - i) / 2] \gets \text{ADCBuf}[l + 1]$
\EndFor
\State $m \gets (m + 1) \mod M$
\State \Return $\text{WaveList}, m$
\end{algorithmic}
\end{algorithm}

The ADCBuf stores the sampling points of both channels in an interleaved format.
Algorithm 1 can be used to transfer a specified range of ADCBuf to the memory space WaveList, separating the two channels: the first half contains the first channel, and the second half contains the second channel.

Given the variability in performance of infrared emitters and receivers due to environmental factors such as time, temperature, and humidity, the baseline voltage across the receiver terminals is not constant.
Periodically updating the trigger threshold values is essential to adapt to to this variability. 
Additionally, establishing an appropriate voltage offset can effectively reduce false triggers caused by voltage fluctuations and dust interference, as described in Algorithm 2.

\begin{algorithm}
\caption{\textbf{AdaptiveThreshold}: Adaptive adjustment of threshold values based on baseline voltage waveform.}
\begin{algorithmic}[1]
\Require Start Position $i \in [0, L-1)$, End Position $j \in (0, L-1]$ such that $0 \leq i < j \leq L-1$, DMA Buffer \( \text{ADCBuf} \in \mathbb{R}^{L} \), Voltage Offset \(\delta_1, \delta_2 \in \mathbb{R}\)
\Ensure Trigger Thresholds $\theta_1, \theta_2 \in \mathbb{R}$
\State $ch1\_sum \gets 0$
\State $ch2\_sum \gets 0$
\For{$l \gets i \text{ to } j \text{ step } 2$}
    \State $ch1\_sum \gets ch1\_sum + ADCBuf[l]$
    \State $ch2\_sum \gets ch2\_sum + ADCBuf[l + 1]$
\EndFor
\State $\theta_1 \gets ch1\_sum / \frac{(j - i)}{2} + \delta_1$
\State $\theta_2 \gets ch2\_sum / \frac{(j - i)}{2} + \delta_2$
\State \Return $\theta_1, \theta_2$
\end{algorithmic}
\end{algorithm}

\begin{algorithm}[!t]
\caption{\textbf{CheckThreshold}: Check for significant jumps within a specified window that exceed the threshold.}
\begin{algorithmic}[1]
\Require Start Index $i \in [0, L-1)$, End Index $j \in (0, L-1]$ such that $0 \leq i < j \leq L-1$, ADC Buffer \( \text{ADCBuf} \in \mathbb{R}^{L} \), Trigger Thresholds $\theta_1, \theta_2 \in \mathbb{R}$, Jump Range $\Delta_1, \Delta_2 \in \mathbb{R}$
\Ensure Trigger Position $p \in [0, L-8]$
\State $j \gets j - 8$
\If{$i < j$}
    \For{$l \gets i \text{ to } j \text{ step } 2$}
        \If{$\text{ADCBuf}[l] \geq \theta_1$}
            \If{$\text{ADCBuf}[l] + \Delta_1 \leq \text{ADCBuf}[l + 8]$ \textbf{or} $\text{ADCBuf}[l] \geq \text{ADCBuf}[l + 8] + \Delta_1$}
                \State $p \gets l$
                \State \textbf{break}
            \EndIf
        \EndIf
        \If{$\text{ADCBuf}[l+1] \geq \theta_2$}
            \If{$\text{ADCBuf}[l+1] + \Delta_2 \leq \text{ADCBuf}[l + 9]$ \textbf{or} $\text{ADCBuf}[l+1] \geq \text{ADCBuf}[l + 9] + \Delta_2$}
                \State $p \gets l + 1$
                \State \textbf{break}
            \EndIf
        \EndIf
    \EndFor
\EndIf
\State \Return $p$
\end{algorithmic}
\end{algorithm}

To prevent pests with strong climbing abilities from clinging to the sensors for extended periods and repeatedly triggering, which could lead to system malfunctions (although such occurrences are infrequent, they can swiftly produce a substantial amount of incorrect data upon happening), we implement an additional condition when checking the threshold.
As shown in Algorithm 3, when the ADC sampling value exceeds the threshold, we check at both ends of an 8-length window for jumps exceeding $\Delta$.
If any channel exhibits such jumps, we return the trigger position; otherwise, we consider continuous triggering by pests and disregard it.

Whenever DMA completes half of a conversion or a full conversion, it triggers a system interrupt. 
In the interrupt handler function, the completed data is processed with \textbf{CheckThreshold}, and if the conditions are met, \textbf{TransferData} is executed. 
In this work, we have set a one-minute interval to perform \textbf{AdaptiveThreshold}.

During the collection process, most perception waveforms are independent, complete, and contain a single pest, as shown in Fig. 5(a.1).
However, there are a few exceptional cases where the likelihood of various abnormal situations varies among different pest species due to their differing habits.
For example, larger size pests like S. z., R. d., and T. c. are more prone to occupying two cycles in a single instance, as depicted in Fig. 5(a.2).
Bore pests like S. z. and R. d. have a tendency to carry grain powder into the monitor, leading to the generation of abnormal waveforms as shown in Fig. 5(a.3).
For pests with strong gripping abilities, such as S. z., may fall together, leading to two pests being detected consecutively, as shown in Fig. 5(a.4).
Considering the aforementioned anomalies,we adopted a small-batch approach for dataset collection.
When a particular PestMoni is triggered more than 100 times, we terminate its monitoring.
For the collected waveforms, we first calculate the sum of all sampling points for each waveform, which is positively correlated with the pest's obstruction of light. 
We classify waveforms with sums that are too low as anomalies caused by voltage fluctuations and discard them.
Next, we check for the presence of consecutive triggered waveforms and merge them if found. 
We then inspect all pest waveforms, extracting those with consecutively triggered (a single waveform with two peaks).
We manually examine the Pest Collection for any obvious grain debris and record the quantity $n$ if found. 
After this, we perform anomaly detection on the remaining waveforms, identifying the $n$ most outlying waveforms, which we attribute to grain debris.
Finally, the remaining waveforms, after the above processing, are classified as pure pest waveforms.

This effort culminated in the development of PestSet, which includes 12,632 waveforms of pest drops, categorized into three distinct types based on the number of pests within a single waveform: zero pests with 140 instances, one pest with 12,389 instances, and two pests with 103 instances.
The single-pest waveforms are further categorized by species as follows: S. z., R. d., T. c., O. s., and C. p., with their respective distributions being 23\%, 21\%, 24\%, 24\%, and 8\%, as depicted in Fig. 1 (central panel).
Fig. 5(b.1) presents waveform samples from PestSet, while Fig. 5(b.2) showcases a statistical analysis (i.e., the distribution of various species across critical features: Duration, Peak, and Energy).
These visualizations demonstrate the comprehensiveness and diversity of PestSet, highlighting its suitability for identification due to the variations among species.
Furthermore, to account for variability between individual PestMonis, we collected standard waveforms from each device.
The procedure involved dropping a black sphere with a diameter of 3.5 mm from the top of Pest Trapping, repeated 100 times for each PestMoni.
This process yielded an extended standard dataset intended for decoupling the data sources, ensuring that the collected data accurately reflects the performance of each individual device.

\begin{figure*}[!t]
\centering
\includegraphics[width=\textwidth]{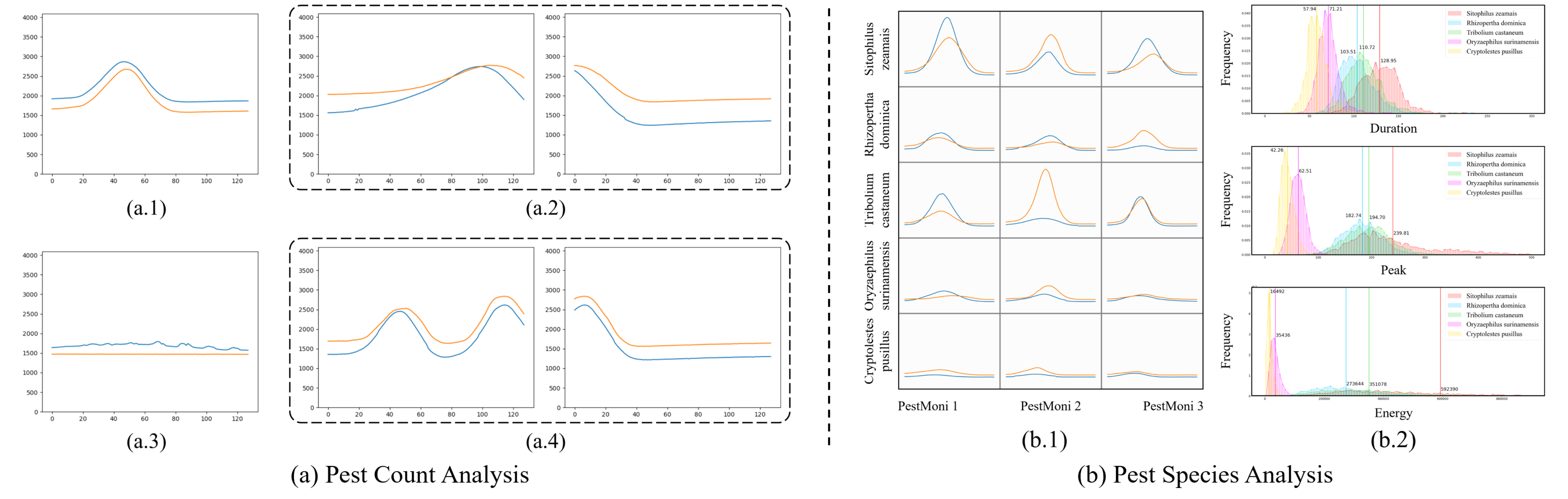}
\caption{Division of Pest Waveforms. 
(a) Based on Pest Count.
(a.1) Normal single pest;
(a.2) Single Pest Spanning Two Cycles;
(a.3) Anomalous Without Pests;
(a.4) Consecutive Drops of Two Pests.
(b) Based on Pest Species. 
(b.1) Waveform samples from three PestMonis, categorized by five species;
(b.2) Frequency distributions of Duration (top), Peak (middle) and Energy (bottom).}
\label{fig5}
\end{figure*}

\subsection{PestFormer Architecture}
As illustrated in Fig. 6 (left), PestFormer is a \textbf{Transformer-based multi-task model specifically designed for the simultaneous tasks of pest counting and species identification} in invisible grain pile storage environments.
While PestMoni is capable of preliminary filtering during data acquisition, unexpected fluctuations may still cause false triggers, interfering with pest counting.
For a single input waveform, denoted as $\boldsymbol{X} \in \mathbb{R}^{T \times C}$, $T$ represents the temporal length and $C$ represents the number of feature channels.
Initially, PestFormer extracts distinctive features from both the time domain and the frequency domain of $\boldsymbol{X}$, including the mean (average signal strength), standard deviation (variation or dispersion), maximum value (peak signal strength), and median (middle value of the signal).
We also consider the first and third quartiles (25th and 75th percentiles), interquartile range (difference between the quartiles), skewness (asymmetry of the signal distribution), kurtosis (tailedness of the distribution), and energy (total power of the signal).
Dimension reduction is then applied to the original waveform and its frequency representation obtained via Fast Fourier Transform.
These selected features and reduced representations are concatenated and processed by a feed-forward network to yield an accurate pest counting output.

For species identification, the task differs significantly from counting.
The \textbf{inter-class differences} among certain pest waveforms are so subtle that manual feature extraction is nearly impossible (see Fig. 5(b.1)). 
Additionally, the heterogeneity of PestMonis used during data acquisition amplifies the \textbf{intra-class differences}, making species identification more challenging. 
Although training a unique model for each device might be a direct solution, it is highly labor-intensive and impractical for real-world applications.
Accurate pest drop location and effective identification require the fusion of data from both sensors. 
However, individual differences introduced during the production of PestMoni result in this fusion being neither uniform nor fixed.

\textbf{Conditional Modification Module.}
In light of these challenges, we propose a Conditional Modification Module (CMM) to refine pest waveform identification and mitigate performance degradation across different PestMonis. 
Specifically, for each PestMoni, we collected waveforms using a black sphere to create a reference waveform pool, denoted as $\mathcal{R} \in \mathbb{R}^{N \times T \times C}$, where $N$ signifies the pool capacity, and $T$ and $C$ are the same as previously defined.
To enhance the generalization ability, we perform random sampling on the reference pool to construct the Ref-waves, as follows:
\begin{align}
\boldsymbol{R} = \text{Samp}(\mathcal{R}, k), \quad k \leq N
\end{align}
where $\text{Samp}(\cdot)$ denotes the random sampling, $\boldsymbol{R}$ represents the Ref-waves, $k$ is the sample size.
Therefore, the identification process could be modified by the Ref-waves to achieve device-aware identification, formulated as follow:
\begin{gather}
\boldsymbol{H}^0 = \text{Norm}\left(\boldsymbol{X}\right) \\
\boldsymbol{H}^{l+1} = \text{CMM}\left(\boldsymbol{H}^{l},\boldsymbol{R}\right) + \boldsymbol{H}^{l}, \quad l=0, \cdots, L-1 \\
\boldsymbol{Y} = \text{FFN}\left(\boldsymbol{H}^{L}\right)
\end{gather}
where, $L$ represents as the number of CMMs, with $l$ as the index.
The input waveform $\boldsymbol{X}$ is initially processed by a Normalizer (Norm) to yield the hidden features $\boldsymbol{H}^0$, which are then fed into the CMM along with the Ref-waves. 
For the $l$-th CMM, Ref-waves undergo a projection to obtain a weight matrix $\boldsymbol{W}$, which is employed for diverse inter-channel fusions of $\boldsymbol{H}$, expanding the data dimensions, as follow:
\begin{gather}
\boldsymbol{W}=\text{Projection}(\boldsymbol{R}) \in \mathbb{R}^{C \times C^{\prime}} \\
\tilde{\boldsymbol{H}}^l=\boldsymbol{H}^{l} \boldsymbol{W} 
\end{gather}

\begin{figure*}[!t]
\centering
\includegraphics[width=0.93\textwidth]{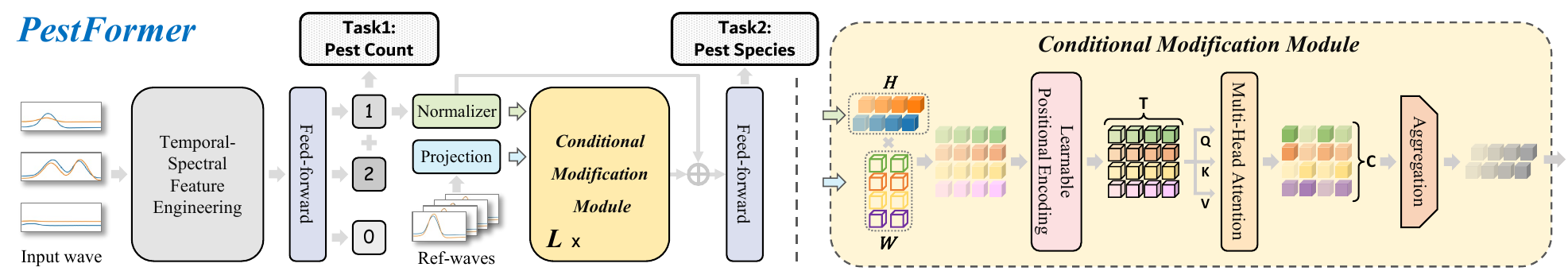}
\caption{Overall architecture of the multi-task PestFormer: The input waveform sequentially undergoes pest counting and species identification.
PestFormer stacks Conditional Modification Modules residually, modifying the input based on Ref-waves.
By integrating multi-head attention and multi-scale aggregation, it achieves lengthwise and channelwise integration, thereby mitigating performance degradation across various PestMonis.}
\label{fig6}
\end{figure*}

Following this, a learnable positional encoding is applied to the expanded data $\tilde{\boldsymbol{H}}^l$ to capture its intrinsic sequential and spatial relationships.
Unlike traditional sinusoidal embedding, this approach can automatically adjust through training alongside the sequential data.
Given an input sequence \(\boldsymbol{H} = [h_1, h_2, ..., h_T]\), where \(h_i\) represents the input at position \(i\), the learnable positional embedding vectors \(\boldsymbol{P} = [p_1, p_2, ..., p_T]\) are a set of embedding vectors with the same length as the input sequence, where \(p_i\) is the embedding vector at position \(i\). 
The final vectors input into the model \(\boldsymbol{E} = [e_1, e_2, ..., e_T]\) are defined as follows:
\begin{gather}
e_i = h_i + p_i
\end{gather}

During the training, the positional embedding vectors \(\boldsymbol{P}\) are optimized along with the other parameters of the model.
Subsequently, the data is processed by a multi-head attention mechanism to capture representations along the temporal dimension. 
This mechanism improves the model's ability to capture various relationships and dependencies in the data.
In multi-head attention, multiple attention heads operate in parallel, each learning different aspects of the input. 
The outputs of these attention heads are concatenated and linearly transformed to produce the final output. 
This process enables the model to consider information from different representation subspaces at different positions.

Given an input sequence of vectors \(\boldsymbol{E} = [e_1, e_2, ..., e_T]\), the multi-head attention mechanism involves:

1. Linear Projections: Projecting the input vectors into queries (\(\boldsymbol{Q}\)), keys (\(\boldsymbol{K}\)), and values (\(\boldsymbol{V}\)) using learned weight matrices \(\boldsymbol{W}_Q\), \(\boldsymbol{W}_K\), and \(\boldsymbol{W}_V\).

\begin{gather}
\boldsymbol{Q} = \boldsymbol{E} \boldsymbol{W}_Q, \quad \boldsymbol{K} = \boldsymbol{E} \boldsymbol{W}_K, \quad \boldsymbol{V} = \boldsymbol{E} \boldsymbol{W}_V
\end{gather}

2. Scaled Dot-Product Attention: Computing scaled dot-product attention for each head:

\begin{gather}
\text{Attention}(\boldsymbol{Q}, \boldsymbol{K}, \boldsymbol{V}) = \text{softmax}\left(\frac{\boldsymbol{Q} \boldsymbol{K}^T}{\sqrt{d_k}}\right) \boldsymbol{V}
\end{gather}

where \(d_k\) is the dimension of the key vectors.

3. Combining Heads: Concatenating the outputs of all heads and projecting them:

\begin{gather}
\text{Attention}(\mathbf{Q}, \mathbf{K}, \mathbf{V}) = \text{Concat}(\text{head}_1, ..., \text{head}_h) \boldsymbol{W}_O
\end{gather}

where \(\text{head}_i = \text{Attention}(\boldsymbol{Q}_i, \boldsymbol{K}_i, \boldsymbol{V}_i)\) and \(\boldsymbol{W}_O\) is the learned weight matrix for the output projection.

Finally, cross-channel aggregation is achieved through multi-scale 2D convolutional kernels, consolidating information across different feature channels.
This dual-dimensional analysis integrates information effectively, enhancing the model's ability to utilize the underlying patterns within the waveform data.
Then, the CMM output is combined with the input and fed to the next CMM layer.
The entire CMM is integrated into the PestFormer through a residual connection.
Finally, the processed variables generate the output $\boldsymbol{Y}$ via a Feed-Forward Network (FFN) realized as a multi-layer perceptron (MLP), thereby achieving pest species identification.

\section{Experiments}
To assess the proposed Pest Manager framework, we conducted a series of targeted experiments evaluating PestMoni's design, PestSet's availability, and PestFormer's performance.
The evaluation includes sensitivity tests for PestMoni, comprehensive verification of PestSet, comparative studies and ablation analysis for PestFormer, and an investigation into the effects of the CMM on enhancing data consistency across devices.
To sum up, these experiments validate the invisible pest monitoring approach in almost all aspects.

\subsection{Rationality of PestMoni}
\textbf{\textit{1) Device Trigger Accuracy.}} 
PestMoni was tested in the laboratory to evaluate its trigger accuracy for monitoring common stored-grain pests.
Initially, 500 pests of each species were evenly introduced into the wheat.
The counting results were checked against the actual number of pests in the Pest Collection once the trigger count exceeded 200.
The absolute value of the difference between the triggered waveform quantity and the actual number of pests, divided by the actual number of pests, was used as the error metric to determine trigger accuracy.
To ensure reliability, three PestMonis were tested simultaneously in an artificially controlled dark environment, as shown in Fig. 4(a), designed to mimic the conditions of an enclosed grain pile storage setting.

\begin{table}[h]
\centering
\caption{Comparison of trigger accuracy for common grain pests across monitors. The symbol $-$ indicates missing data. Standard deviations are in the lower right corner.}
\resizebox{0.95\linewidth}{!}{
\begin{tabular}{lccccc}
\toprule
\multirow{2}{*}{\vspace{-0.15cm}Monitors} & \multicolumn{5}{c}{Stored Grain Pests} \\
\cmidrule(lr){2-6}
& S. z. /  S. o. (\%) & R. d. (\%) & T. c. (\%) & O. s. (\%) & C. p. (\%) \\
\midrule
OITD-PIS \cite{b27} & $88.6$ & $69.9$ & $76.4$ & $-$ & $-$ \\
EGPIC \cite{b6} & $83_{\pm 8}$ & $92_{\pm 6}$ & $85_{\pm 1}$ & $91_{\pm 6}$ & $-$ \\
Our PestMoni & $98.22_{\pm 1.23}$ & $98.83_{\pm 1.11}$ & $99.20_{\pm 0.24}$ & $99.28_{\pm 0.77}$ & $99.12_{\pm 0.80}$ \\
\bottomrule
\end{tabular}}
\end{table}

The test results of PestMoni are presented in Table II, along with performance comparisons to two other previous monitors.
Notably, Sitophilus zeamais (S. z.) and Sitophilus oryzae (S. o.) are similar in size and morphology, making them difficult to distinguish, even manually, and thus they are grouped together.
PestMoni demonstrates significantly higher trigger accuracy for five different pest species compared to other devices. 
Remarkably, even for the smallest pest C. p., which is typically the most easily overlooked, the detection accuracy has reached over 99\%.
This performance highlights the advanced sensitivity of the proposed PestMoni, which can handle pests of diverse sizes.

\vspace{2mm}

\textbf{\textit{2) Asymmetric Diode Layout and Tailored Circuit Design.}} 
Three combinations of PestMoni designs were tested to validate the enhancement of perception sensitivity: a symmetrical layout with traditional circuit, an asymmetrical layout with traditional circuit, and an asymmetrical layout with tailored circuit.
For the evaluation experiments, we selected C. p., the most difficult-to-detect pest.
Each combination was tested by manually dropping pests 10 times per individual configuration.
\textbf{Notably, we propose a metric to measure the overall sensitivity.}
This metric is the simplified harmonic mean of the peak voltage changes in two channels, calculated as follows:

\begin{align}
\eta = \frac{\Delta V_{\text{ch1}} \cdot \Delta V_{\text{ch2}}}{\Delta V_{\text{ch1}} + \Delta V_{\text{ch2}}}
\end{align} 
where $\Delta V$ (unit: mV) represents the peak voltage change of a single channel.
The metric $\eta$ represents the overall sensitivity of two channels, strongly influenced by extremely low values. 
A higher $\eta$ indicates stronger perception capabilities and a more balanced sensitivity between channels.

\begin{table}[h]
\centering
\caption{Analysis of the Effectiveness of Asymmetric Diode Layout and Tailored Circuit Design.}
\resizebox{0.95\linewidth}{!}{
\begin{tabular}{lcccc}
\toprule
\multirow{2}{*}{\vspace{-0.15cm}Combinations} & \multicolumn{4}{c}{Experimental Groups} \\
\cmidrule(lr){2-5}
& PestMoni 1 & PestMoni 2 & PestMoni 3 & Average \\
\midrule
Symm.  + Conv. & $37.60_{\pm 19.41}$ & $47.56_{\pm 18.08}$ & $47.17_{\pm 20.44}$ & $44.11_{\pm 19.87}$ \\
Asymm. + Conv. & $56.09_{\pm 19.43}$ & $64.02_{\pm 22.87}$ & $55.28_{\pm 20.78}$ & $58.46_{\pm 21.44}$ \\
Asymm. + Tail. & $81.33_{\pm 15.76}$ & $83.44_{\pm 14.41}$ & $83.04_{\pm 17.37}$ & $82.61_{\pm 15.92}$ \\
\bottomrule
\end{tabular}
}
\end{table}

As shown in Table III, $\eta$ significantly outperforms symmetric diode layouts across all devices employing asymmetric diode layouts.
This superiority underscores the effectiveness of the proposed design in enhancing perceptual sensitivity in both channels.
Additionally, the incorporation of tailored circuit not only further boosts perceptual sensitivity but also minimizes variability in performance across multiple units.
Such enhancements are pivotal for the robust pest counting and species identification based on PestMoni.

\begin{figure}[h]
\centerline{\includegraphics[width=0.9\columnwidth]{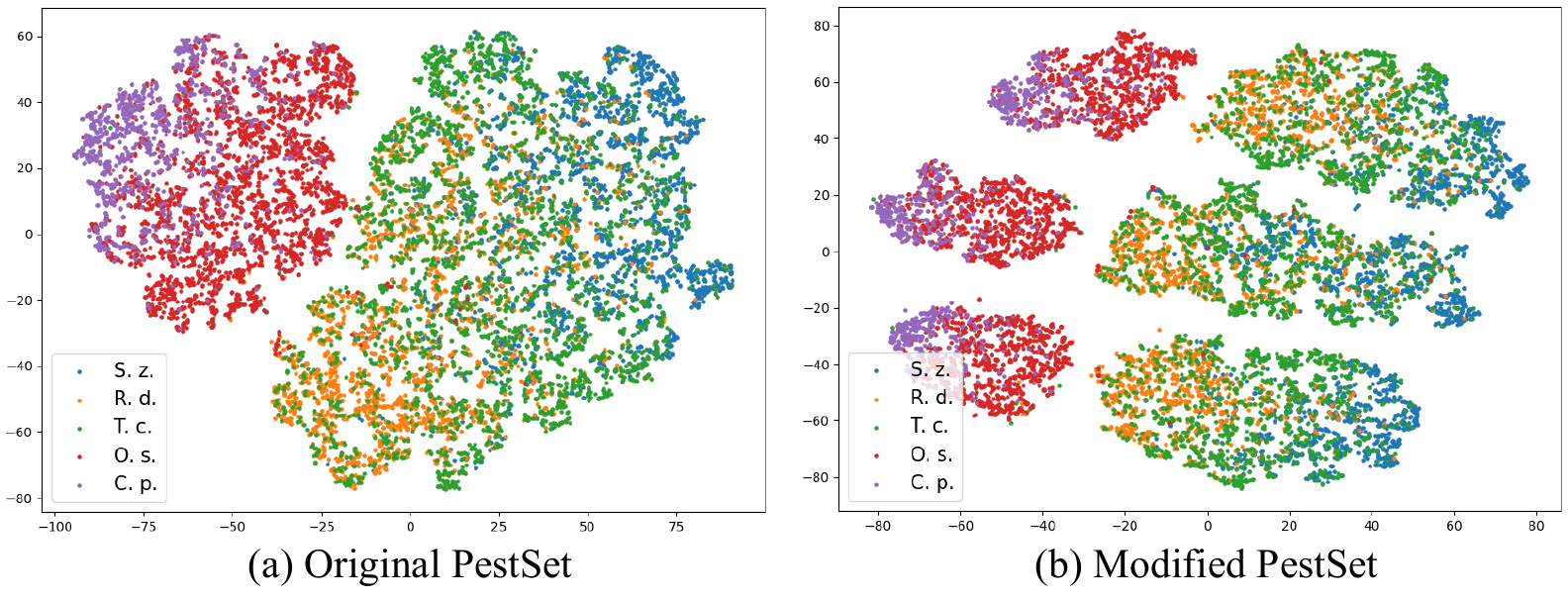}}
\caption{Visualization of pest species distribution using t-SNE with parameters (learning\_rate=200, n\_iter=1000): (a) Original PestSet (b) PestSet modified by PestFormer using Ref-waves.}
\label{fig7}
\end{figure}

\subsection{Availability of PestSet}
To evaluate PestSet, we used t-SNE \cite{b28} to visualize the differences between species in PestSet. 
As shown in Fig. 7(a), the original PestSet exhibits a clear boundary between large-sized pests (S. z., R. d., T. c.) and small-sized pests (O. s., C. p.). 
However, the three large-sized pest species are intermixed, posing a significant challenge for species identification tasks.

To enhance the availability of PestSet, we proposed a modification method using Ref-waves on the original dataset. 
As shown in Fig. 7(b), the modified data points are separated into three groups, effectively decoupling data from different PestMonis.
This process also results in more distinct color bands, which increases the separability between different categories. 
However, despite these improvements, the boundaries between different species remain blurred, necessitating the use of a powerful model to achieve accurate species identification. Thus, we propose PestFormer.

\begin{table*}[!t]

\centering
\caption{Comparison of Time Series Classification Algorithms}
\renewcommand{\arraystretch}{0.9}
\setlength{\aboverulesep}{1pt}
\setlength{\belowrulesep}{2.5pt}
\resizebox{0.9\linewidth}{!}{
\begin{tabular}{lccc|ccccc}
\toprule
\multirow{2.5}{*}{Algorithms} & \multicolumn{3}{c}{Pest Counting} & \multicolumn{5}{c}{Pest Species Identification} \\
\cmidrule(lr){2-4} \cmidrule(lr){5-9}
& Recall & Precision & Accuracy & Kappa & Recall & Precision & F1 Score & Accuracy \\
\midrule
LSTM (1997) & 0.9866 & 0.9871 & 0.9868 & 0.7278 & 0.7667 & 0.7808 & 0.7726 & 0.7877 \\
LSTNet (2018) & 0.7886 & 0.7951 & 0.7895 & 0.3336 & 0.4303 & 0.6052 & 0.3908 & 0.4899 \\
\midrule
TCN (2018) & 0.7046 & 0.7725 & 0.7014 & 0.2961 & 0.3877 & 0.2747 & 0.2981 & 0.4621 \\
TimesNet (2023) & 0.9910 & 0.9912 & 0.9911 & 0.7648 & 0.7891 & 0.8111 & 0.7975 & 0.8168 \\
\midrule
LightTS (2022) & 0.9881 & 0.9885 & 0.9883 & 0.6913 & 0.7379 & 0.7461 & 0.7412 & 0.7591 \\
DLinear (2023) & 0.9356 & 0.9360 & 0.9357 & 0.6546 & 0.7117 & 0.7247 & 0.7172 & 0.7308 \\
\midrule
iTransformer (2024) & 0.9863 & 0.9869 & 0.9865 & 0.7584 & 0.7837 & 0.8009 & 0.7895 & 0.8115 \\
Reformer (2020) & 0.9905 & 0.9908 & 0.9907 & 0.7720 & 0.8012 & 0.8134 & 0.8056 & 0.8220 \\
Transformer (2017) & 0.9910 & 0.9912 & 0.9911 & 0.7984 & 0.8173 & 0.8272 & 0.8205 & 0.8426 \\
Informer (2021) & 0.9903 & 0.9906 & 0.9904 & 0.8203 & 0.8331 & 0.8520 & 0.8407 & 0.8600 \\
\midrule 
PestFormer (Ours) & \textbf{0.9927} & \textbf{0.9929} & \textbf{0.9929} & \textbf{0.8330} & \textbf{0.8525} & \textbf{0.8642} & \textbf{0.8575} & \textbf{0.8697} \\
\bottomrule
\end{tabular}}

\end{table*}

\begin{figure}[!t]
\centerline{\includegraphics[width=0.95\columnwidth]{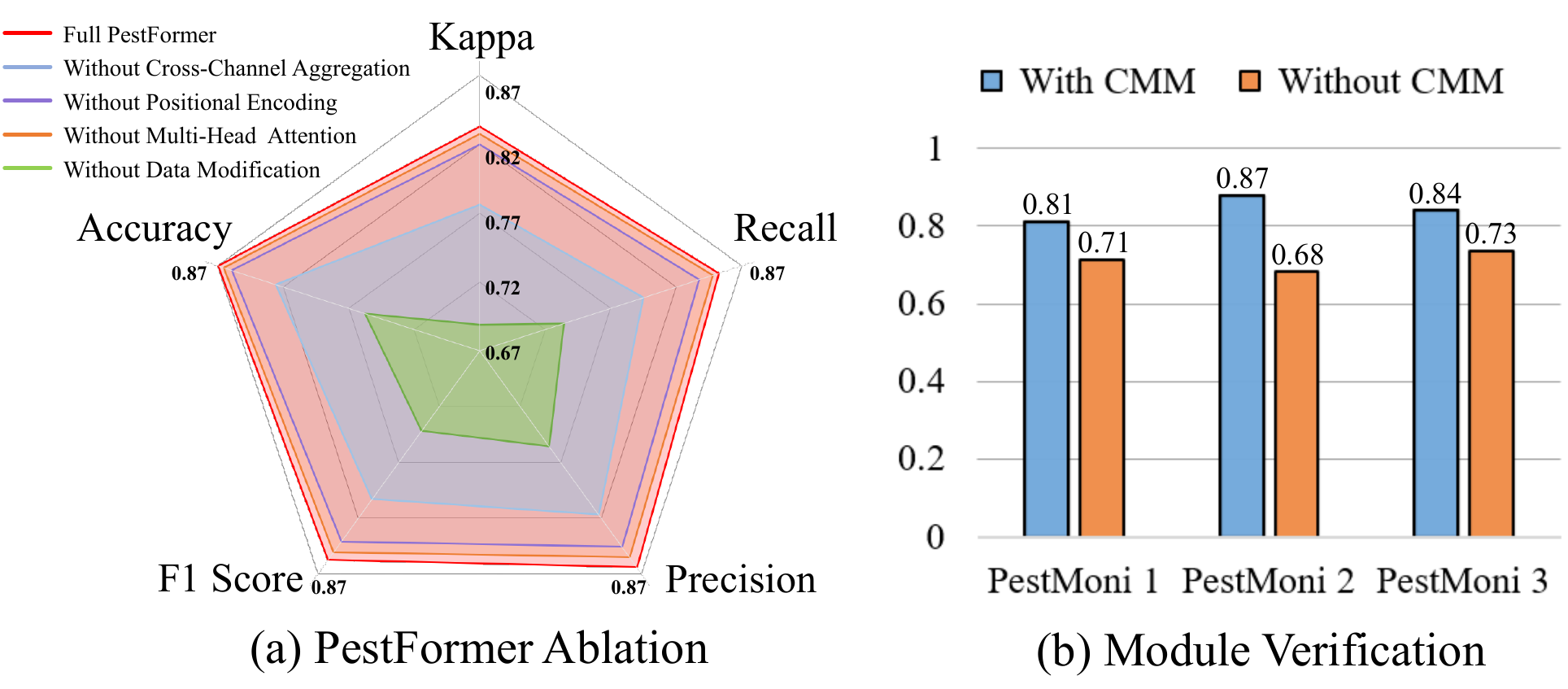}}
\caption{(a) Ablation analysis of PestFormer: Contribution of each component to model performance. (b) CMM module validation: effects of CMM used across different data sources.}
\label{fig8}
\end{figure}

\subsection{Effectiveness of PestFormer}
\textbf{\textit{1) Comparison with TSC Methods.}} We compare PestFormer with several well-acknowledged and advanced TSC methods to evaluate the classification performance, including the RNN-based models: LSTM \cite{b10} and LSTNet \cite{b11}; CNN-based Model: TCN \cite{b25} and TimesNet \cite{b22}; MLP-based models: LightTS \cite{b29} and DLinear \cite{b30}; Transformer-based models: iTransformer \cite{b25}, Reformer \cite{b24}, Transformer \cite{b15} and Informer \cite{b23}.
We have divided the dataset into training, validation, and test sets with a ratio of 6:2:2.
Notably, to address class imbalance in the pest count dataset, we have applied oversampling for equal representation prior the splitting. 
All experiments are conducted with an NVIDIA Tesla V100 with 32GB of memory.
The Adam optimizer has been used with an initial learning rate of 0.001, batch size of 16.
As shown in Table IV, in the Pest Counting task where distinguishing features are clear, most models perform well, with our PestFormer achieving the highest performance. 
In pest species identification, transformer-based models, benefiting from the strong feature extraction capabilities of the attention mechanism, generally excel. Among these, our PestFormer leverages the CMM to achieve the highest classification performance.

\textbf{\textit{2) Ablation Analysis of PestFormer.}}
To better understand the contributions of different components, we conducted ablation studies on PestFormer.
These components are Conditional Modification, Learnable Positional Encoding, Multi-Head Attention, and Average Aggregation.
Each variant is trained and evaluated under the same conditions as the original model, with metrics like accuracy, precision, recall, F1 score, and Kappa to gauge the effects of each modification.
Fig. 8(a) illustrates the pivotal role of each component. 
Eliminating any component leads to reduced performance, with \textbf{Conditional Modification being particularly critical}.
This underscores the importance of conditional modification and its powerful strength in information capture.

\textbf{\textit{3) The Validation of the CMM.}}
Comparative experiments have been conducted on different PestMoni to assess the impact of CMM across diverse data sources.
Initially, the dataset is split into training and testing sets with a 6:4 ratio. 
Subsequently, the testing set is organized by data source.
We train PestFormer with and without CMM using a combined training set. 
The accuracy results from single-dataset testing, presented in Fig. 8(b), demonstrate that PestFormer with CMM significantly outperforms its counterpart without CMM across all devices.
This highlights the robustness and adaptability of PestFormer with CMM, making it a versatile and effective solution for pest monitoring and classification tasks across diverse environments.

\section{Conclusion}
In this paper, we propose Pest Manager, a systematic framework for precise pest counting and identification in invisible grain pile storage environment. 
Specifically, Pest Manager includes an improved grain probe trap PestMoni with two specific designs, a pest drop infrared perception dataset PestSet and a multi-task Transformer-based architecture PestFormer. 
Extensive experiments are conducted to verify the rationality of the sensor design, the availability of the developed dataset, and the effectiveness of our PestFormer. 
As a result, our Pest Manager is a systematic solution for pest management in invisible environment.
However, there are still some shortcomings, such as the device stability in production process, the generalization of algorithms, and limited quantity of dataset. 
In the future, more work is needed to improve the stability and reliability of the entire system. 
Moreover, real-scene system validation also needs to be addressed urgently. 

\bibliographystyle{IEEEtran}
\bibliography{main.bib}

\end{document}